\documentclass{an}
\usepackage{times}
\usepackage{fancyhdr}
\usepackage{graphicx}
\graphicspath{{./fig/}{./png/}}
\newcommand{\EQ}{\begin{equation}}
\newcommand{\EN}{\end{equation}}
\newcommand{\EQA}{\begin{eqnarray}}
\newcommand{\ENA}{\end{eqnarray}}
\newcommand{\eq}[1]{(\ref{#1})}
\newcommand{\Eq}[1]{Eq.~(\ref{#1})}

\newcommand{\Eqss}[2]{Eqs.~(\ref{#1})--(\ref{#2})}
\newcommand{\eqs}[2]{(\ref{#1}) and~(\ref{#2})}
\newcommand{\Sec}[1]{Sec.~\ref{#1}}
\newcommand{\Fig}[1]{Fig.~\ref{#1}}

\newcommand{\Tab}[1]{Table~\ref{#1}}
\newcommand{\Figs}[2]{Figs~\ref{#1} and \ref{#2}}

\newcommand{\bra}[1]{\langle #1\rangle}

\newcommand{\uu}{\mbox{\boldmath $u$} {}}

\newcommand{\grav}{\mbox{\boldmath $g$} {}}
\newcommand{\nab}{\mbox{\boldmath $\nabla$} {}}

\newcommand{\SSSS}{\mbox{\boldmath ${\sf S}$} {}}

\def\half{{\textstyle{1\over2}}}

\def\twothird{{\textstyle{2\over3}}}

\newcommand{\g}{\,{\rm g}}
\newcommand{\s}{\,{\rm s}}

\newcommand{\cm}{\,{\rm cm}}

\newcommand{\kms}{\,{\rm km/s}}

\newcommand{\erg}{\,{\rm erg}}

\newcommand{\yapj}[3]{: #1, {ApJ} {#2}, #3}
\newcommand{\yapjl}[3]{: #1, {ApJ} {#2}, #3}

\newcommand{\yana}[3]{: #1, {A\&A} {#2}, #3}

\newcommand{\yjfm}[3]{: #1, {JFM} {#2}, #3}

\newcommand{\ymn}[3]{: #1, {MNRAS} {#2}, #3}

\newcommand{\yjour}[4]{: #1, {#2}, {#3}, #4}

\newcommand{\ybook}[3]{: #1, {#2} (#3)}

\title{Effect of the radiative background flux in convection}
\author{A. Brandenburg\inst{1} \and K. L. Chan\inst{2}
\and {\AA}. Nordlund\inst{3} \and R. F. Stein\inst{4}}
\institute{
NORDITA, Blegdamsvej 17, DK-2100 Copenhagen \O, Denmark
\and Department of Mathematics, The Hong Kong University of Science \& Technology, Hong Kong
\and Astronomical Observatory / NBIfAFG, Juliane Maries Vej 30, DK-2100 Copenhagen \O, Denmark
\and Department of Physics and Astronomy, Michigan State University, East Lansing, MI 48824, USA}

\abstract{
Numerical simulations of turbulent stratified convection are used to study
models with approximately the same convective flux, but different
radiative fluxes. 
As the radiative flux is decreased, for constant 
convective flux: the entropy jump at the top of the convection zone 
becomes steeper, the temperature fluctuations increase and the 
velocity fluctuations decrease in magnitude, and the distance that
low entropy fluid from the surface can penetrate increases.  Velocity 
and temperature fluctuations follow mixing length scaling laws.
}

\date{Received 15 July 2005; accepted 16 August 2005; published online 1 September 2005}
\begin{document}
\maketitle

\section{Introduction}

In modeling stellar convection it is important to make the models
resemble the stars as much as possible.
A major difficulty in producing realistic simulations of deep
stellar convection is the large ratio between thermal and dynamical
time scales. This is because in dynamical units the solar thermal flux 
($F_\odot=7\times10^{10}\erg\cm^{-2}\s^{-1}$) is very small, 
\EQ
{F_\odot\over\rho c_{\rm s}^3}\sim4\times10^{-11},
\EN
at the bottom of the convection zone (where
$\rho=0.2\g\cm^{-3}$ and $c_{\rm s}=200\kms$).
At the surface
this ratio is of order $10^{-1}$, but already two megameters below the
surface the ratio is $10^{-3}$.
This ratio is basically equal
to the ratio of the dynamical to thermal time scales. 

For a common class of models studied
by many workers (we will refer to them as polytropic models), most of
the energy is carried by radiation instead of convection. 
This corresponds to the polytropic index $m$
of the associated hydro-thermal equilibrium solution having the value
$m=1$.  In this case flow characteristics are different from
those in the convection dominated transport regime, even for large
Rayleigh numbers. In the range $-1<m\ll1$, however,
most of the flux is carried by convection. 

In order for simulations of convection to be representative of late type
stars the ratio of radiative to total flux must be small in the 
convection zone.
This has prompted Chan \&
Sofia (1986, 1989) to study models where the radiative flux is removed
entirely and is replaced by a subgrid scale flux.
Unlike the radiative flux, which is proportional to the
temperature gradient, the subgrid scale or small-scale eddy flux is
proportional to the entropy gradient (e.g.\ R\"udiger 1989).
Others have chosen to model the surface
layers and the granulation (Nordlund 1982; Stein \& Nordlund 1989, 1998;
Kim \& Chan 1998, Robinson et al.\ 2003, V\"ogler et al.\ 2005).
In those models radiation is only important near the
surface layers and practically absent beneath the surface, although 
diffusive energy flux is still necessary for numerical simulations to 
be stable.

The aim of this paper is to explore the effect of varying the diffusive
radiative flux while keeping the convective flux in the convection
zone approximately constant. From mixing length arguments one would
expect that for negligible radiative flux the turbulent velocities, 
temperature fluctuations and other
dynamical aspects of the convection only depend on the magnitude of the
convective flux.  However, this
approximation breaks down once the radiative flux is no
longer very small compared to the convective flux.

We consider models using piecewise
polytropic layers. Such models have been widely studied and
they possess the advantage that the properties of stable and unstable
layers are easier to manipulate than in models with the more realistic
Kramers' opacities, for example, where the radiative diffusivity depends
on density and temperature rather than just on depth.
However, we also include a comparison of realistic solar models with
the same convective flux but varying diffusive subgrid scale energy flux
in the interior and hence varying (turbulent) Prandtl number.

There are two main results of our investigation.  First, the properties
of the convection depend on the ratio of radiative to convective flux
when the radiative flux is not negligible.  Increasing radiative energy
diffusion reduces the temperature fluctuations which requires larger
velocities to carry the same convective flux.  Increasing radiative
energy diffusion also raises the entropy of the downdrafts and inhibits
their descent.  
Second, as the resolution increases the dependence of
convective properties on the Prandtl number decreases. However, at low resolutions
the dependence is different when the viscosity is altered than when the
radiative conductivity is altered.  Increasing the radiative
conductivity to lower the Prandtl number has the effects described
above.  Decreasing the viscosity to lower the Prandtl number
(increasing the Reynolds number) enhances small scale structures and
increases the turbulence.

We first discuss the setup of our model, the equations that are solved 
and how we change the fractional
radiative flux in the convection zone while leaving everything else
unchanged (\Sec{Seqn}).
The polytropic model results are given in Sects~\ref{Sres}
and \ref{Ssubgrid}, the realistic solar simulation results are presented in
\Sec{Srealist}, and our conclusions are given in \Sec{Sconcl}.

\section{The model}
\label{Seqn}

\subsection{Equations}

In our polytropic models we solve
the continuity, momentum, and energy equations in
non-conservative form,
\EQ
{D\ln\rho\over Dt}=-\nab\cdot\uu,
\label{dlnrhodt}
\EN
\EQ
{D\uu\over Dt}=-(\gamma-1)\,(e\nab\ln\rho+\nab e)
+\grav+{1\over\rho}\nab\cdot(2\nu\rho\SSSS),
\EN
\EQ
{D e\over Dt}=-(\gamma-1)\,e\nab\cdot\uu
+{1\over\rho}\nab\cdot{K}\nab T
+2\nu\SSSS^2-{e-e_0\over\tau_e(z)},
\label{dedt}
\EN
where ${\sf S}_{ij}=\half(\partial_j u_i+\partial_i u_j
-\twothird\delta_{ij}\partial_k u_k)$ is the (traceless) strain tensor,
$\nu=\mbox{const}$ is the kinematic viscosity.
We assume a perfect gas, so the pressure is given by
\EQ
p=(\gamma-1)\rho e,
\EN
where $\rho$ is density, $e=c_{\rm v} T$ is the internal energy, $T$ is
temperature, and $c_{\rm v}$ is the specific heat at constant volume.

Below the photosphere the radiative diffusion approximation is valid
so the vertical component of the radiative flux is 
\EQ
F_{\rm rad}=K dT/dz, 
\EN
where $K$ is the radiative conductivity, $T$ is temperature, and $z$ is depth
(increasing downward).
We impose a cooling layer at the surface where
$\tau_e(z)$ is a cooling time,
and $e_0$ is the reference value of the
internal energy $e$ at the top of the layer.
The value of $e_0$ is quantified by a parameter
$\xi_0=(\gamma-1)e_0/(gd)\equiv H_p^{\rm(top)}/d$, which is the
pressure scale height at the top of the box divided by the depth of
the unstable layer, $d$.  The pressure scale height is, 
$H_p=(d\ln p/dz)^{-1}={\cal R}T/(\mu g)$, 
where ${\cal R}$ is the universal
gas constant and $\mu$ the mean molecular weight. 

We adopt the basic setup of the model of Brandenburg et al.\ (1996,
hereafter BJNRST; see also Hurlburt et al.\ 1986)
where the vertical profile of ${\cal K}(z)\equiv K/c_{\rm v}$
is assumed such that ${\cal K}$ is piecewise constant in
three different layers,
${\cal K}={\cal K}_1$ in $z_1\leq z\leq z_2$,
${\cal K}={\cal K}_2$ in $z_2\leq z\leq z_3$, and
${\cal K}={\cal K}_3$ in $z_3\leq z\leq z_4$. In our case
$z_1=-0.15$, $z_2=0$, $z_3=1$, and $z_4=2.8$.

\subsection{Radiative conductivity}

In this paper we want to study the effects of varying the radiative
flux in the convection zone, $z_2\leq z\leq z_3$, so we have to vary
the value of ${\cal K}_2$. In the following we discuss the relation
between ${\cal K}_2$ and the anticipated radiative flux $F_{\rm rad}$
in the convection zone, which will be a good approximation to the actual
radiative flux obtained by solving \Eqss{dlnrhodt}{dedt}.

In astrophysics the radiative flux is often written in the form
\EQ
F_{\rm rad}
=KT\,\left({d\ln p\over dz}\right)\,\left({d\ln T\over d\ln p}\right)
={KT\over H_p}\,\nabla = {Kg\over{\cal R}/\mu}\,\nabla.
\EN
Here 
\EQ
\nabla=d\ln T/d\ln p 
\EN
characterizes the temperature stratification. 
We can define a radiative gradient, 
$\nabla_{\rm rad}$, that would be necessary to transport all the flux 
radiatively,
\EQ
F_{\rm tot}={Kg\over{\cal R}/\mu}\,\nabla_{\rm rad}.
\label{eq:Ftot}
\EN
In the deep radiative interior of the sun we have $F_{\rm rad}=F_{\rm tot}$,
and therefore $\nabla=\nabla_{\rm rad}$. 
In most of the convection zone the actual stratification is close
to adiabatic, so $\nabla\approx\nabla_{\rm ad}\equiv1-\gamma^{-1}<\nabla_{\rm rad}$, 
since $F_{\rm rad}<F_{\rm tot}$.

To set up a polytropic model it is customary to specify ${\cal K}$
in terms of the polytropic index $m$ (e.g., Hurlburt \& Toomre 1984),
instead of $\nabla_{\rm rad}$. If all the flux were carried by radiation
we would have $p(z)\sim T^{m+1}$, and so $\nabla_{\rm rad}=(m+1)^{-1}$.
The value of the radiative conductivity is expressed in terms of the
polytropic index from \Eq{eq:Ftot} as
\EQ
{\cal K}_i = {K_i\over{c_v}} = {F_{\rm tot}\over g}\,(\gamma-1)(m_i+1).
\EN
The three values of ${\cal K}_i$ ($i=1,2,3$) are given in terms
of polytropic indices $m_i$ via the above equation.
In all cases reported below we use $m_3=3$ and vary the value of $m_2$. 
Near the top there is cooling ($\tau_e\neq0$ in $z_1<z<z_2$,
and $\tau_e\rightarrow0$ for $z>z_2$). Therefore, almost all the flux
in this layer is transported by the corresponding cooling flux and
the diffusive radiative flux in the uppermost layer is insignificant.
Hence, the value of $m_1$ does not affect the
stratification.  In most of the cases we put $m_1=-0.9$. 

The fractional radiative flux in the convection zone is
\EQ
{F_{\rm rad}\over F_{\rm tot}}\approx{\nabla_{\rm ad}\over\nabla_{\rm rad}}
=\left(1-{1\over\gamma}\right)(m+1),
\;\;\mbox{for}\;\; m<{1\over\gamma-1},
\label{eq:fractional1}
\EN
where we have assumed $\nabla\approx\nabla_{\rm ad}$.
In all those models where $m=1$ (e.g.\ Hurlburt \& Toomre 1984, 1986;
Cattaneo et al.\ 1990, 1991; Brandenburg et al.\ 1990, 1996; Brummell
et al.\ 1996, Ossendrijver et al.\ 2002, Ziegler \& R\"udiger 2003,
K\"apyl\"a et al.\ 2004)
and $\gamma=5/3$, the fractional radiative flux is as
large as 80\%. In order to study polytropic models with
$F_{\rm rad}\ll F_{\rm tot}$ we must approach the limit $m\rightarrow-1$.
In this paper we calculate a series of such models starting with $m=1$
(the standard case considered in many papers) down to $m=-0.9$.
As the value of $m$ is lowered, smaller fractional radiative
fluxes are obtained. 
In the upper layers of the solar convection zone
this ratio is close to the ratio of the mean free path of the photons
to the pressure scale height and can be as small as $10^{-5}$. In deeper
layers of the solar convection zone the radiative flux becomes progressively
more important and reaches 50\% of the total flux at 0.75 solar radii.

It may seem unphysical to talk about negative values of $m$, because
the density would increase with height, but such a `top-heavy' arrangement
only means that the stratification is then very unstable.
Of course, the pressure still decreases with height, because
$m+1$ is positive. Negative values of $m$ (but with $m>-1$) result from
a very poor radiative conductivity, which is indeed quite common in the
outer layers of all late-type stars.

It should be emphasized that $m$ only characterizes the {\it associated}
thermal equilibrium hydrostatic solution, which is of course unstable
for $m<m_{\rm ad}\equiv(\gamma-1)^{-1}=3/2$. In that case convection
develops, making the stratification close to adiabatic. Thus, the {\it
effective} value of $m$ will then always be close to the marginal value
$m_{\rm ad}$. The significance of the $m$ used here (as opposed to the
effective $m$) is that it gives a nondimensional measure of the
fractional radiative flux.  

We carry out a parameter survey by varying the values of $m_2$ and 
hence ${\cal K_2}$, which corresponds to varying $\nabla_{\rm rad}$ in the
convection zone and the total flux $F_{\rm tot}$. In practice we
fix the value of ${F}_{\rm conv}$ and calculate the total flux
${F}$ from ${F}_{\rm conv}$ for a given value of $m_2$.
In \Tab{Tfluxes} we give the fractional
fluxes for some values of $m$, using \Eq{eq:fractional1}. Here we have
assumed $F_{\rm tot}=F_{\rm rad}+F_{\rm conv}$, i.e.\ we have neglected
kinetic and viscous fluxes which are of course included in the
simulations.
\begin{table}[tb]\caption{
Fractional radiative and convective fluxes for a few values of $m$,
as obtained from \Eq{eq:fractional1}.
}\vspace{12pt}\centerline{\begin{tabular}{ccc}
$m$ & $F_{\rm rad}/F_{\rm tot}$ & $F_{\rm conv}/F_{\rm tot}$ \\
\hline
$ 1$   & 80\% & 20\% \\
$ 0$   & 40\% & 60\% \\
$-0.5$ & 20\% & 80\% \\
$-0.8$ &  8\% & 92\% \\
$-0.9$ &  4\% & 96\% \\
\label{Tfluxes}\end{tabular}}\end{table}

\subsection{Nondimensional quantities}

Nondimensional quantities are adopted by measuring length in units
of $d$, time in units of $\sqrt{d/g}$ and $\rho$ in units of its initial
value, $\rho_0$, at $z=z_3$, i.e.\ at the bottom of the convection zone.
In all cases we use a box size of $L_x=L_y=4$ with $z_4=2.8$.
The flux is then expressed in terms of the non-dimensional quantity
\EQ
{\cal F}={F_{\rm tot}\over\rho_0\,(gd)^{3/2}},
\EN
In the same units we define
\EQ
{\cal F}_{\rm conv}=
\left[1-\left(1-{1\over\gamma}\right)(m+1)\right]{\cal F},
\EN
which is a measure of the convective flux. (The actual convective flux,
$F_{\rm conv}$, is of course not a constant, but varies with height,
and reaches a maximum of around ${\cal F}_{\rm conv}$.) 

The value of
$\nu$ is taken to be as small as possible for a given mesh resolution.
For $50^3$ we can typically use $\nu=5\times10^{-3}$ (in units of
$\sqrt{gd^3}$). For a resolution of $200^3$ we were able to lower
the viscosity to $\nu=2.4\times10^{-4}$.

The quantity ${\cal F}$ determines the `mean' thermal diffusivity,
$\overline{\chi}_3={\cal K}_3/(\gamma\rho_0)$, via
\EQ
\overline{\chi}_i
=\sqrt{gd^3}\,{\cal F}\,\nabla_{\rm ad}/\nabla_{\rm rad}^{(i)},
\EN
where $\nabla_{\rm rad}^{(i)}$ is the radiative gradient in layer $i$.
(Note that in this approach the actual thermal diffusivity decreases with
depth in each of the three layers.) The nondimensional flux can also be
related to the Rayleigh number,
\EQ
\mbox{Ra}
={gd^4\over\nu\overline{\chi}_2 c_p}\left({ds\over dz}\right)_0,
\EN
which characterizes the degree of instability of the hydrostatic solution
in the middle of the unstable layer, $i=2$. Here,
\EQ
{d\over c_p}\left({ds\over dz}\right)_0
={1-\nabla_{\rm ad}/\nabla_{\rm rad}^{(2)}\over
\half+\xi_0/\nabla_{\rm rad}^{(2)}}
\EN
is the normalized entropy gradient of the associated hydrostatic
solution for the same ${\cal K}(z)$ profile (see BJNRST).
We can then express $\mbox{Ra}$ in terms of ${\cal F}$ as
\EQ
\mbox{Ra}={1\over\mbox{Pr}\,{\cal F}^2}\,
\left({\nabla_{\rm rad}^{(2)}\over\nabla_{\rm ad}}\right)^2
{1-\nabla_{\rm ad}/\nabla_{\rm rad}^{(2)}\over
\half+\xi_0/\nabla_{\rm rad}^{(2)}},
\EN
where $\mbox{Pr}=\nu/\overline{\chi}_2$ is the Prandtl number.
In the astrophysically interesting limit,
$\nabla_{\rm rad}^{(2)}\rightarrow\infty$, we have
\EQ
\mbox{Ra}\rightarrow\mbox{Ra}^*=2\mbox{Pr}^{-1}\,{\cal F}^{-2}\,
(\nabla_{\rm rad}^{(2)}/\nabla_{\rm ad})^2,
\EN
so large Rayleigh numbers correspond to small normalized fluxes and hence small
velocities. This is at first glance somewhat surprising, because large
Rayleigh numbers are normally associated with more vigorous convection
and therefore large velocities.  However, while the velocity decreases
in absolute units, it does increase relative to viscosity and radiative
diffusivity, i.e.\ the Reynolds and Peclet numbers do indeed increase.

\subsection{The initial condition}
\label{Sinit}

The ratio of radiative to convective flux can easily be decreased within the
framework of standard polytropic models, provided the polytropic index $m$
is close to $-1$. In that case, however, it is a bad idea to use
the corresponding polytropic solution as the initial condition, because
it is extremely far away from the final solution and would take a very
long time to relax to the final state. Instead, we use
a simplified mixing length model with an empirically determined free
parameter such that the entropy at the bottom of the convection zone and
the radiative interior beneath is close to that finally obtained in the
actual simulation.\footnote{The reason why we chose to specify our simulations
in terms of polytropic indices is mainly that it makes good contact
with previous approaches using polytropic solutions. It is important to
realize, however, that specifying the value of $m$ in the convection
zone is really just a way of specifying the nondimensional radiative
conductivity.}

An initial condition that yields a solution on almost the right
adiabat is obtained by assuming that the entropy is not constant,
but that the negative entropy gradient is a function of the convective
heat flux. According to mixing length theory the superadiabatic
gradient in the convection zone is
\EQ
\nabla-\nabla_{\rm ad}=k\,({F_{\rm conv}/\rho c_{\rm s}^3})^{2/3},
\EN
where $k$ is a nondimensional coefficient (related to the mixing length
alpha parameter), which we determined empirically for our models to be
$k=1.5$.

In terms of $e$ and $\ln\rho$, our initial condition can then be written as
\EQ
de/dz=m_{\rm ad} g\,\nabla,
\label{e_init}
\EN
\EQ
d\ln\rho/dz=m_{\rm ad} g\,(1-\nabla)/e,
\label{r_init}
\EN
where
\EQA
\nabla=\left\{
\begin{array}{ll}
0 & \quad\mbox{if}\quad z_1\leq z\leq z_2\\
\nabla_{\rm ad}+k\,({F_{\rm conv}/\rho c_{\rm s}^3})^{2/3}
  &\quad\mbox{if}\quad z_2 < z < z_3\\
\nabla_{\rm rad}^{(3)}=(m_3+1)^{-1}
 & \quad\mbox{if}\quad z_3\leq z\leq z_4
\end{array}
\right.
\ENA
and varies smoothly between the three layers. The initial stratification is
obtained by integrating Eqs~\eq{e_init} and \eq{r_init} from $z=z_1$
downwards to $z_4$. The starting value of $\rho$ at $z=z_1$ is adjusted
iteratively such that $\rho(z_3)=\rho_0=1$, i.e.\ the density at
the bottom of the convection zone is unity. This is especially useful
if the extent of the box is changed, because this would otherwise affect
the density in all layers.

\begin{figure}[t!]\includegraphics[width=\columnwidth]{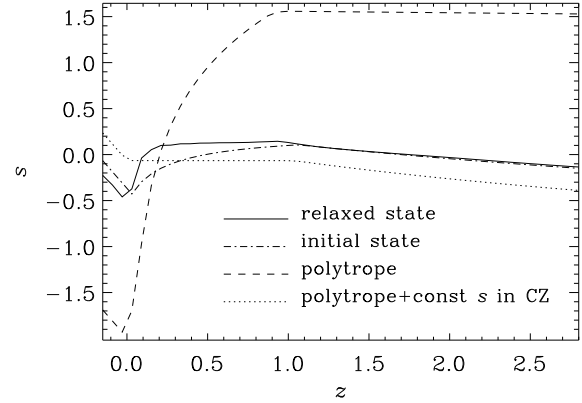}\caption[]{
Entropy in the final and initial states, compared with the
corresponding polytropic model, as well as a polytropic model
with constant entropy within the convection zone (CZ).
${\cal F}_{\rm conv}=0.01$.
}\label{Fpss_poly}\end{figure}

In \Fig{Fpss_poly} we compare the entropy obtained from the initial
condition as derived above with that obtained from the actual simulations.
We also compare with the entropy derived from the associated 
polytropic hydrostatic
thermal equilibrium solution to show just how far away from the final
state that solution would be. Finally, we also compare with a solution
where the entropy was assumed constant within the convection zone,
which is better than the associated hydrostatic thermal equilibrium,
but still with the wrong entropy at the bottom of the convection zone.

As in previous work the initial velocity is random. This is realized
by a superposition of randomly located spherical blobs of radius 0.1,
where the velocity points in random directions.
 
\subsection{Boundary conditions}

In the horizontal directions we adopt periodic boundary conditions
and at the top and bottom we adopt impenetrable, stress-free boundary
conditions, i.e.\
\EQ
{\partial u_x\over\partial z}={\partial u_y\over\partial z}=u_z=0
\quad\mbox{at}\quad z=z_1,z_4.
\EN
The boundaries are sufficiently far away so the 
boundary conditions do not significantly affect the flow properties in the
convection zone proper, which is the layer $z_2\leq z\leq z_3$.
Because of the cooling term in \Eq{dedt}
the value of $e$ at the top is always close to the reference value
$e_0$ which, in turn, is proportional to $\xi_0$ and to the
pressure scale height at the top. Thus, decreasing the value of $\xi_0$ leads
to stronger driving of convection and to stronger stratification in the
top layer. However, $\xi_0$ cannot be decreased too much for a
given numerical resolution, and so we chose $\xi_0=0.2$ in all cases.
In the top layer $z_1<z<z_2$ the inverse cooling time $\tau_e^{-1}$ is
10 and goes smoothly to zero for $z>z_2$.

\section{Results}
\label{Sres}
\subsection{Models}

We have carried out a series of calculations all with the same value
of ${\cal F}_{\rm conv}$ and varying values of $m_2$ (hereafter referred
to as $m$). In a first series of models we used $\nu=6\times10^{-3}$.
The corresponding values of Prandtl and Rayleigh numbers are given in
Table~\ref{Tfluxes2}.

\begin{table}[tb]\caption{
Parameters for a model with ${\cal F}_{\rm conv}=0.01$, $\xi_0=0.2$,
$\nu=6\times10^{-3}$ and different values of $m$.
}\vspace{12pt}\centerline{\begin{tabular}{ccccc}
$m$ & ${\cal F}_{\rm tot}$ & ${\cal F}_{\rm rad}$ & Pr & Ra \\
\hline
$ 1$   & 0.0500 & 0.0400 &0.15 & $9.3\times10^2$ \\
$ 0$   & 0.0167 & 0.0067 & 0.9 & $2.1\times10^4$ \\
$-0.5$ & 0.0125 & 0.0025 & 2.4 & $8.9\times10^4$ \\
$-0.8$ & 0.0109 & 0.0009 & 6.9 & $3.3\times10^5$ \\
$-0.9$ & 0.0104 & 0.0004 &  14 & $7.4\times10^5$ \\
\label{Tfluxes2}\end{tabular}}\end{table}

Note that the Prandtl number is no longer small, except in the case
$m=1$. However, for $m=1$ and ${\cal F}_{\rm conv}=0.01$ the Rayleigh
number is already so small that the instability to convection is
suppressed. Therefore we have also studied another series of models
where the convective flux is reduced by a factor of 5; see
Table~\ref{Tfluxes3}. Here however we have only calculated the cases
$m=0$ and 1, because otherwise the radiative diffusivity would become
so small that the resolution would be insufficient.

\begin{table}[tb]\caption{
Parameters for a model with ${\cal F}_{\rm conv}=0.002$ and
$\nu=5\times10^{-3}$ and different values of $m$.
}\vspace{12pt}\centerline{\begin{tabular}{ccccc}
$m$ & ${\cal F}_{\rm tot}$ & ${\cal F}_{\rm rad}$ & Pr & Ra \\
\hline
$ 1.0$ & 0.01000 & 0.00800 & 0.6 & $5.6\times10^3$ \\
$ 0.0$ & 0.00333 & 0.00133 & 3.8 & $1.3\times10^5$ \\
\label{Tfluxes3}\end{tabular}}\end{table}

Finally, we consider a series of models with varying Prandtl number
and constant convective flux.
Lowering the radiative flux automatically increases the
Prandtl number. In the sun the Prandtl number is less than one, which
was no longer the case in models with low radiative flux. We therefore
also studied the effects of lowering the Prandtl number by lowering the
viscosity. In \Tab{Tfluxes4} we give the parameters for a series of models
where Pr is varied by changing both $\nu$ and $\overline{\chi}_2$.
The question
is whether or not certain properties of convection continue to depend on
Prandtl number as the turbulence becomes more vigorous (small viscosity
and radiative diffusivity).  For instance, we expect that for
sufficiently large Reynolds number the large scale flow properties will
no longer depend on the microscopic values of viscosity and radiative
diffusivity.

\begin{table}[tb]\caption{
Parameters for a model with ${\cal F}_{\rm conv}=0.01$ and
different values of $m$. $\nu$ and $\overline{\chi}_2$ are given
in units of $10^{-4}$, i.e.\ $\nu_{-4}=\nu/10^{-4}$ and
$\chi_{-4}=\overline{\chi}_2/10^{-4}$. The mesh resolution
is also given. The asterisk in the last column indicates that
the duration of the run was short and not yet fully relaxed, so the data
may not be reliable.
}\vspace{12pt}\centerline{\begin{tabular}{ccccccccc}
$m$    & $\nu_{-4}$
             & $\chi_{-4}$
                   & Pr  &     Ra      & resol.& $k_u$& $k_T$\\
\hline
$-0.5$ & 60  &  25 & 2.4 &$9\times10^4$&$ 50^3$& 0.34 & 1.13 &   \\
$-0.5$ & 12  &  25 & 0.5 &$4\times10^5$&$100^3$& 0.46 & 1.07 &   \\
$-0.9$ & 60  &  4  & 14  &$7\times10^5$&$100^3$& 0.29 & 1.25 &   \\
$-0.9$ & 12  &  4  & 2.9 &$4\times10^6$&$100^3$& 0.41 & 1.14 &   \\
$-0.9$ & 2.4 &  4  & 0.6 &$2\times10^7$&$200^3$& 0.48 & 1.14 &$\!\!$*$\!\!$\\
$-.98$ & 12  & 0.8 & 15  &$2\times10^7$&$100^3$& 0.38 & 1.22 &   \\
$-.98$ & 2.4 & 0.8 & 3.0 &$1\times10^8$&$200^3$& 0.43 & 1.16 &   \\
\label{Tfluxes4}\end{tabular}}\end{table}

\subsection{Entropy stratification}

\begin{figure}[t!]\includegraphics[width=\columnwidth]{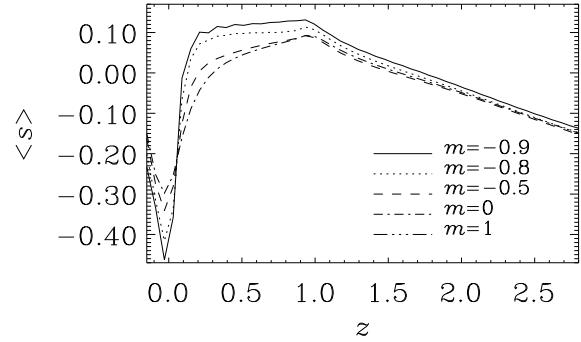}\caption[]{
Comparison of the horizontally averaged
entropy stratification for different values of $m$ and
${\cal F}_{\rm conv}=0.01$.
}\label{Fcompare_ssmean}\end{figure}

As the radiative flux is lowered, the mean entropy in the convection
zone becomes more nearly constant and the superadiabatic gradient at
the top becomes steeper; see \Fig{Fcompare_ssmean}. The mean
entropy beneath the convection zone is only slightly affected.

\begin{figure}[t!]\includegraphics[width=.9\columnwidth]{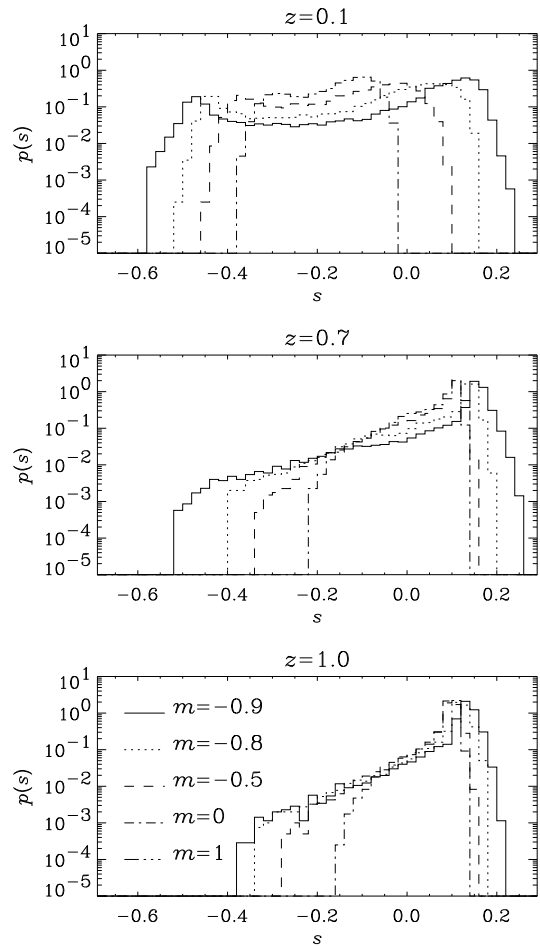}\caption[]{
Entropy histograms for different values of $m$ and ${\cal F}_{\rm conv}=0.01$
at three different values of $z$.
}\label{Fchk_entro_all}\end{figure}

\begin{figure}[t!]\includegraphics[width=\columnwidth]{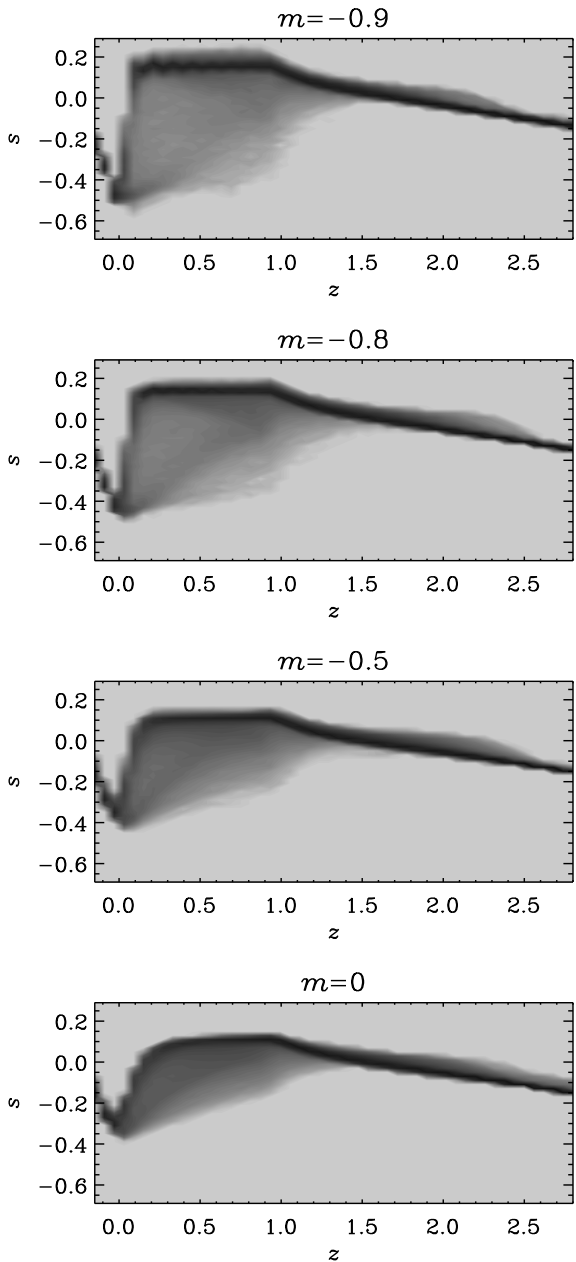}\caption[]{
Entropy histograms for different values of $m$ and ${\cal F}_{\rm conv}=0.01$.
}\label{Fchk_entro_grey}\end{figure}

\begin{figure}[t!]\includegraphics[width=\columnwidth]{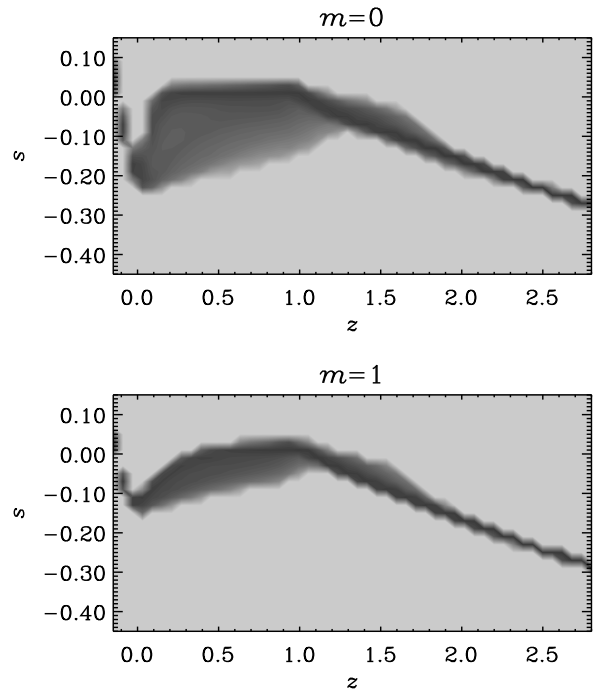}\caption[]{
Entropy histograms for different values of $m$ and ${\cal F}_{\rm conv}=0.002$.
Note that the range on the ordinate is reduced compared to the previous figure.
}\label{Fchk_entro_grey_f}\end{figure}

As the radiative flux decreases the minimum entropy in the interior of
the convection zone decreases 
(\Fig{Fchk_entro_all} and \Fig{Fchk_entro_grey})
and approaches the minimum value that occurs at the top of the 
convection zone. These low entropy elements are only
produced at the top where cooling is important. As the 
diffusive energy exchange decreases
at least some fluid elements make it all the way
from the top to the bottom with very little mixing or heating. This is
clearly a consequence of the reduced radiative diffusivity in the
cases with low radiative flux. In the case $m=1$ the deviations from the
median of the entropy are rather small; see \Fig{Fchk_entro_grey_f},
where ${\cal F}_{\rm conv}=0.002$ instead of ${\cal F}_{\rm conv}=0.01$,
so the entropy drop at the surface is obviously smaller.

\begin{figure}[t!]\includegraphics[width=\columnwidth]{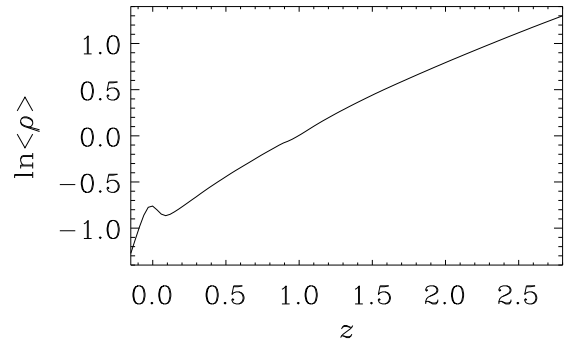}\caption[]{
Horizontally averaged density stratification for $m=-0.9$ and
${\cal F}_{\rm conv}=0.01$.
}\label{Fcompare_rhom}\end{figure}

As the radiative flux is lowered, the drop of entropy and temperature
at the surface becomes more sudden. This is because with less radiative
diffusivity (i.e.\ with less radiative energy transfer) the thermal
boundary layer at the top becomes thinner.
At the same time the pressure must fall off smoothly, because the
pressure is primarily determined by approximate hydrostatic balance.
This can cause a rather pronounced density inversion near the
top, as is seen in \Fig{Fcompare_rhom}, which is occasionally also
seen in the more realistic solar simulations.

\subsection{Convective and kinetic flux profiles}

\begin{figure}[t!]\includegraphics[width=\columnwidth]{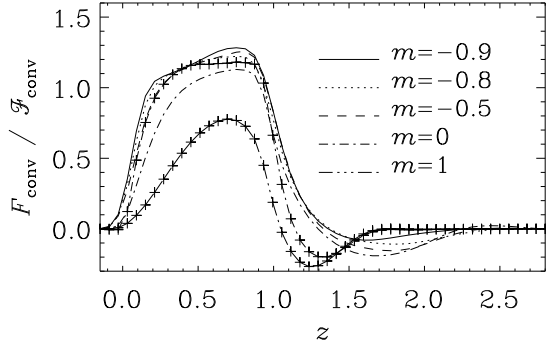}\caption[]{
Comparison of the convective flux for different values of $m$
and ${\cal F}_{\rm conv}=0.01$ or 0.002 (marked by additional + symbols).
}\label{Fcompare_fconv}\end{figure}

\begin{figure}[t!]\includegraphics[width=\columnwidth]{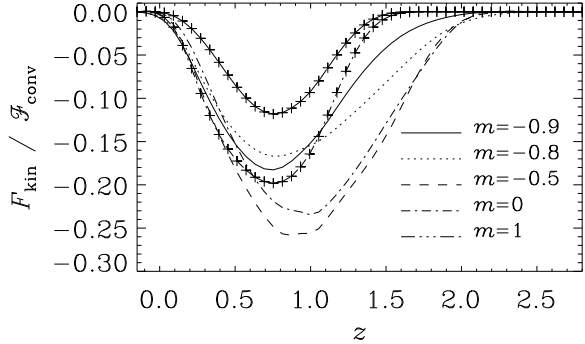}\caption[]{
Comparison of the kinetic energy flux for different values of $m$
and ${\cal F}_{\rm conv}=0.01$ or 0.002 (marked by additional + symbols).
}\label{Fcompare_fkin}\end{figure}

The mean convective flux profiles (\Fig{Fcompare_fconv}) begin to
converge to one and the same profile as the radiative flux is lowered.
The differences between $m=-0.8$ and $-0.9$ are small, suggesting
that with $m=-0.8$ or $-0.9$ the convective properties of the
simulations (convective velocities and temperature fluctuations) begin
to converge. However, the convective flux is not constant in the
convection zone.  This is because of some contribution of the kinetic
energy flux, which is plotted separately in \Fig{Fcompare_fkin}. As we lower
the convective flux (from ${\cal F}_{\rm conv}=0.01$ to ${\cal F}_{\rm
conv}=0.002$), the convective velocities decrease and therefore also
the kinetic flux.

The depth of the overshoot layer is clearly reduced as we decrease the
convective flux (Figs \ref{Fcompare_fconv} and \ref{Fcompare_fkin}).
This is in qualitative agreement with theories linking
the depth of the overshoot layer to the magnitude of the convective
velocities and hence to the convective flux (e.g.\ Hurlburt et
al.\ 1994; Singh et al.\ 1998).

\subsection{Velocity and temperature fluctuations}
\label{SSveltempflucts}

\begin{figure}[t!]\includegraphics[width=\columnwidth]{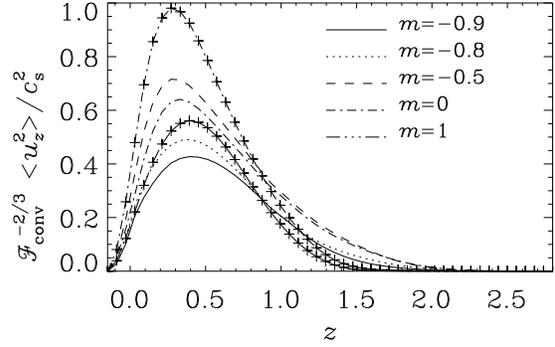}\caption[]{
Comparison of the normalized vertical velocity fluctuations for
different values of $m$ and ${\cal F}_{\rm conv}=0.01$ or 0.002
(marked by additional + symbols). The normalization factor is
${\cal F}_{\rm conv}^{-2/3}$.
}\label{Fcompare_turms1}\end{figure}

\begin{figure}[t!]\includegraphics[width=\columnwidth]{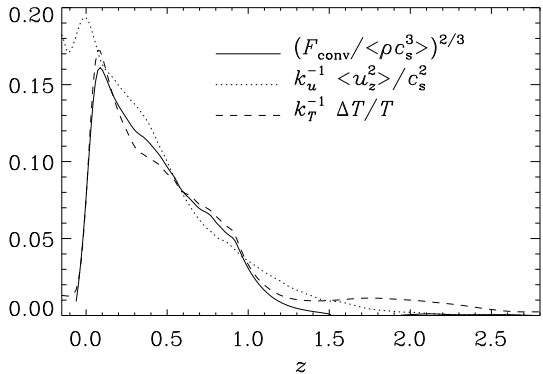}\caption[]{
Vertical profiles of the normalized mean squared vertical velocity
fluctuations and temperature fluctuations, compared with the normalized
convective flux raised to the power $2/3$. Note the good agreement
between the three curves within the convection zone proper.
}\label{Fprfluctn}\end{figure}

According to mixing length theory, the vertical velocity variance
is proportional to the relative temperature fluctuation, so
$\bra{u_z^2}\sim(\Delta T/T)\,g\ell$, where $\ell$ is the mixing
length, and $g\ell\sim c_{\rm s}^2$. The magnitude of velocity and
temperature fluctuations (denoted below by primes) is determined
by the convective flux via
\EQ
F_{\rm conv}=\bra{(\rho u_z)'(c_p T)'}
\sim\bra{\rho}\bra{u_z^2}^{1/2}c_p\Delta T.
\EN
This estimate implies that, apart from factors of order unity
(to be determined below),
\EQ
{\Delta T\over T}\sim{\bra{u_z^2}\over c_{\rm s}^2}
\sim\left({F_{\rm conv}\over\rho c_{\rm s}^3}\right)^{2/3}.
\label{Tufluct}
\EN
In \Figs{Fcompare_turms1}{Fcompare_turms2} we show the normalized
velocity and temperature fluctuations for different runs: the profiles are
basically similar. Except for the cases $m=1$ and $m=0$, the magnitude of
the velocity fluctuations decreases and the magnitude of the temperature
fluctuations increases as the radiative flux and/or the value of ${\cal
F}_{\rm conv}$ is lowered. Also, the ratios in \Eq{Tufluct} are indeed
of order unity. It turns out that this is not only valid globally,
but also locally; see \Fig{Fprfluctn}. Here we have defined the coefficients
\EQ
k_u={\bra{\bra{u_z^2}/c_{\rm s}^2}_{\rm CZ}\over
\bra{F_{\rm conv}/(\rho c_{\rm s}^3)}^{2/3}},\quad
k_T={\bra{\Delta T/T}\over\bra{F_{\rm conv}/(\rho c_{\rm s}^3)}^{2/3}},
\label{ku_kT_def}
\EN
where the averages are taken over the convection zone proper, i.e.\
in $z_2<z<z_3$.

\begin{figure}[t!]\includegraphics[width=\columnwidth]{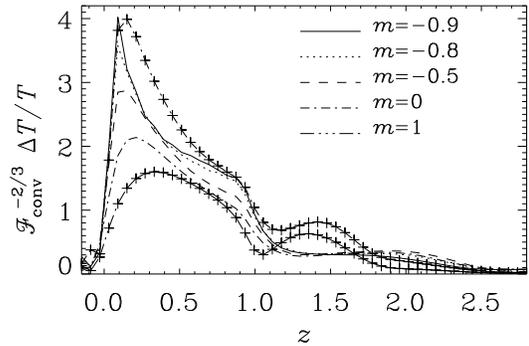}\caption[]{
Comparison of the normalized temperature fluctuations for different values
of $m$ and ${\cal F}_{\rm conv}=0.01$ or 0.002 (marked by additional +
symbols). The normalization factor is ${\cal F}_{\rm conv}^{-2/3}$.
}\label{Fcompare_turms2}\end{figure}

\begin{figure}[t!]\includegraphics[width=\columnwidth]{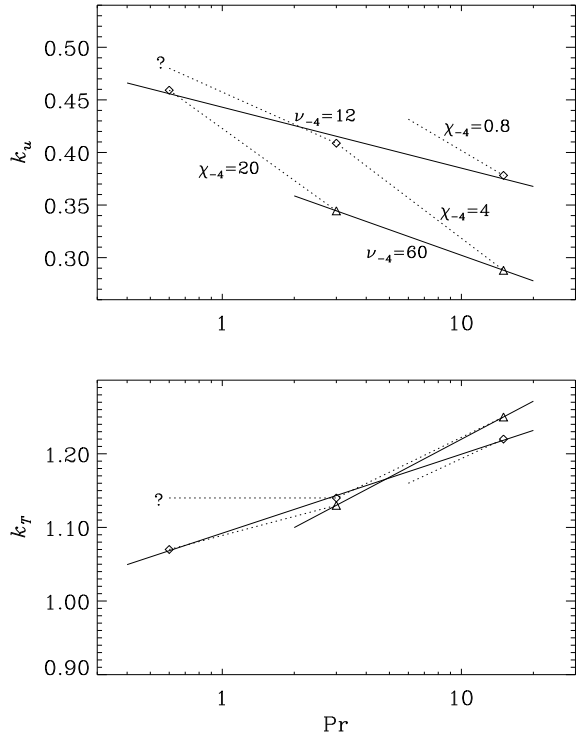}\caption[]{
Dependence of $k_u$ and $k_T$ on Pr. Note that $\nu$ is constant along
solid lines and $\overline{\chi}_2$ and hence $m$ are constant along
dotted lines. $\nu$ and $\overline{\chi}_2$ are given in units of
$10^{-4}$, i.e.\ $\nu_{-4}=\nu/10^{-4}$ and $\nu_{-4}=\nu/10^{-4}$.
}\label{Fpconver}\end{figure}

The dependence of these scaling relations \eqs{Tufluct}{ku_kT_def} on Prandtl
number is shown as a plot of $k_u$ and $k_T$ on Pr in \Fig{Fpconver}.
The dependence of $k_T$ on Pr is more or less unique,
independent of whether the change in Pr is accomplished by changing $\nu$
or $\overline{\chi}_2$. By contrast, the dependence of $k_u$ on Pr is
not unique. For constant values of $\nu$, $k_u$ is nearly constant,
while it increases with decreasing values of $\overline{\chi}_2$.
We explain this result as follows.

As $\overline{\chi}_2$ increases (Pr decreases), temperature fluctuations
are smoothed out, thus decreasing $\Delta T/T$. This would decrease the
convective flux (which is proportional to the product of temperature and
vertical velocity fluctuations), but since the convective flux is kept
constant this can only be achieved by increasing the vertical velocity
fluctuations and hence $k_u$.

The fact that $k_u$ and $k_T$ remain Pr-dependent even at reasonably large
Reynolds numbers is somewhat surprising. Our largest Reynolds number,
based on the rms velocity in the convection zone proper and the depth
of the unstable layer, is around 1000. It is possible that one needs to
go to much larger values before $k_u$ and $k_T$ become independent of Pr.

\subsection{Morphology} 

\begin{figure}[t!]\includegraphics[width=\columnwidth]{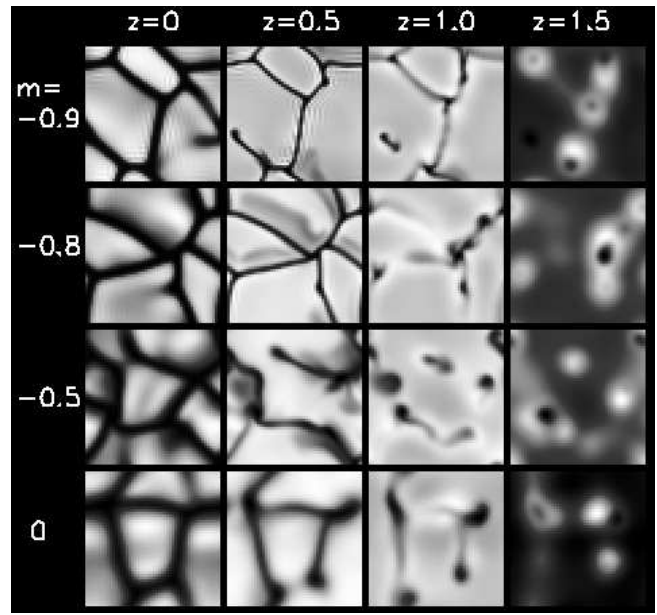}\caption[]{
Horizontal slices of temperature at 4 different levels for four
different values of $m$. ${\cal F}_{\rm conv}=0.01$.
}\label{FMMc2}\end{figure}

\Fig{FMMc2} shows temperature (or $e$) in horizontal planes at
various depths. At the top of the unstable layer the temperature displays
a familiar granular pattern with cool downdraft lanes. As $m$
approaches $-0.9$ the pattern becomes generally sharper and more
complex and of smaller scale. At the bottom of the convection zone
($z=1$) there are isolated cool downdrafts. For $m=-0.9$ the granular
surface pattern still prevails, but this is connected with
weak density stratification. As the downdrafts enter the lower
overshoot layer ($z=1.5$) they become warmer. This is now due to the
exterior entropy being lower than that of the downdrafts, which carry
entropy from the upper convection zone.

\begin{figure}[t!]\includegraphics[width=\columnwidth]{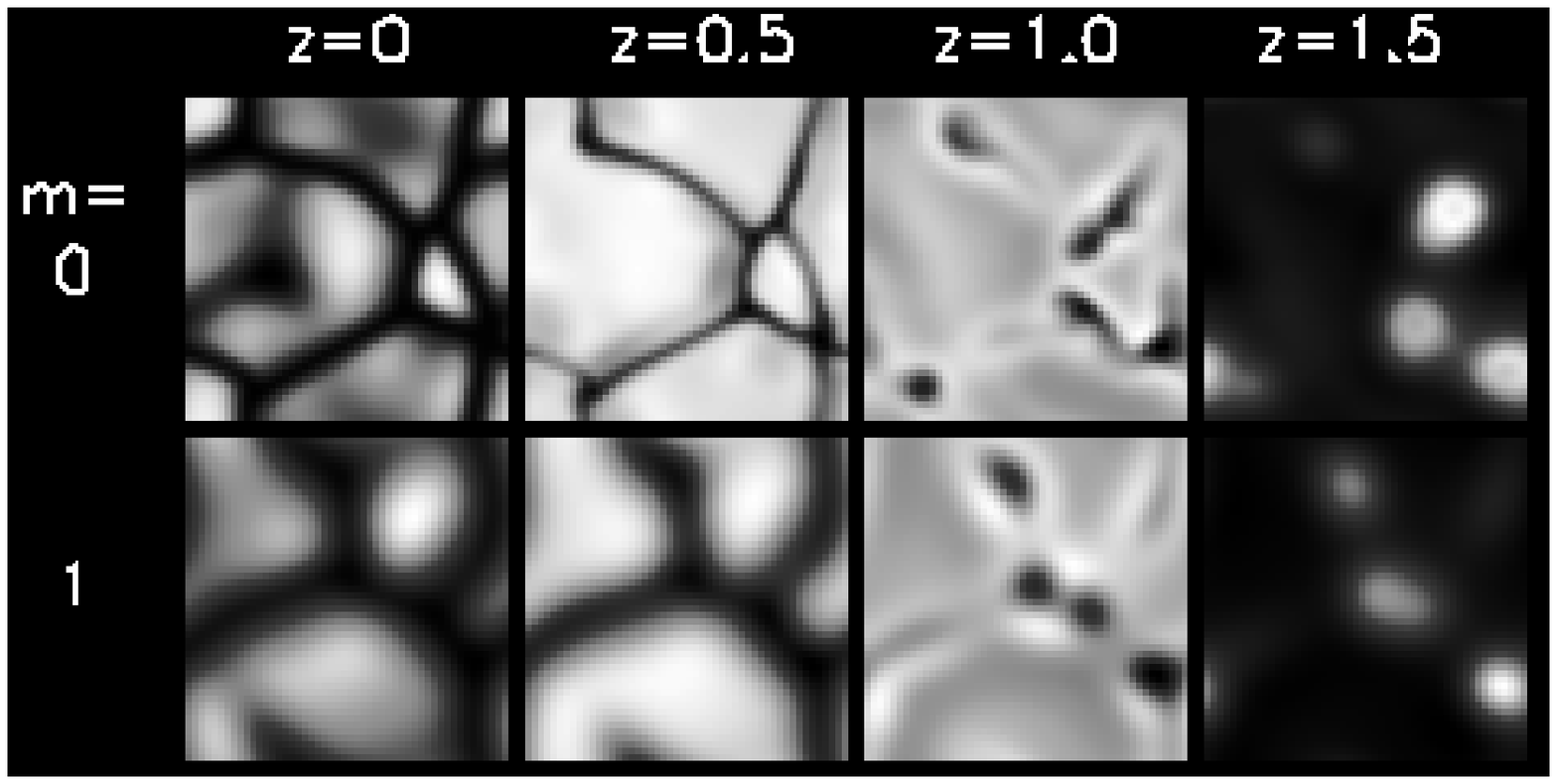}\caption[]{
Horizontal slices of temperature at 4 different levels for two
different values of $m$. ${\cal F}_{\rm conv}=0.002$.
}\label{FMMf}\end{figure}

\begin{figure}[t!]\includegraphics[width=\columnwidth]{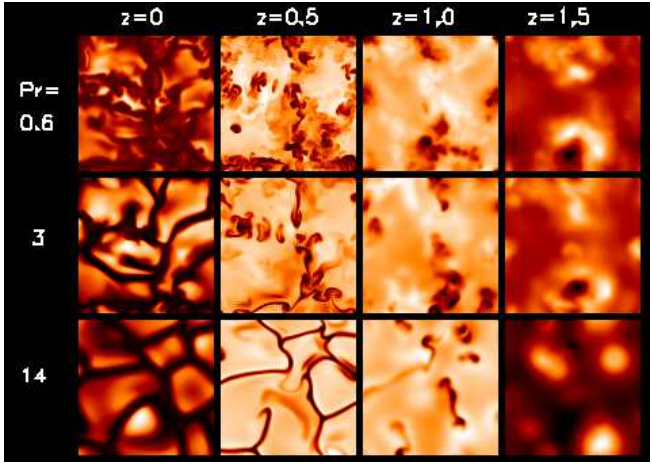}\caption[]{
Horizontal slices of temperature at 4 different levels for $m=-0.9$ and
three different values of $\nu$. ${\cal F}_{\rm conv}=0.01$. A resolution
of $200^3$ meshpoints was used for Pr=0.6, and $100^3$ for Pr=3 and 14.
}\label{FMM_hires}\end{figure}

When the convective flux is decreased (${\cal F}_{\rm conv}=0.002$; see
\Fig{FMMf}), the temperature pattern becomes sharper again, but the
structure and the typical number of cells remains about the same. 
On the other hand, if $\nu$ is lowered the temperature becomes much more
intermittent and of significantly smaller scale; see \Fig{FMM_hires}.

\section{Subgrid scale models}
\label{Ssubgrid}

\begin{figure}[t!]\includegraphics[width=\columnwidth]{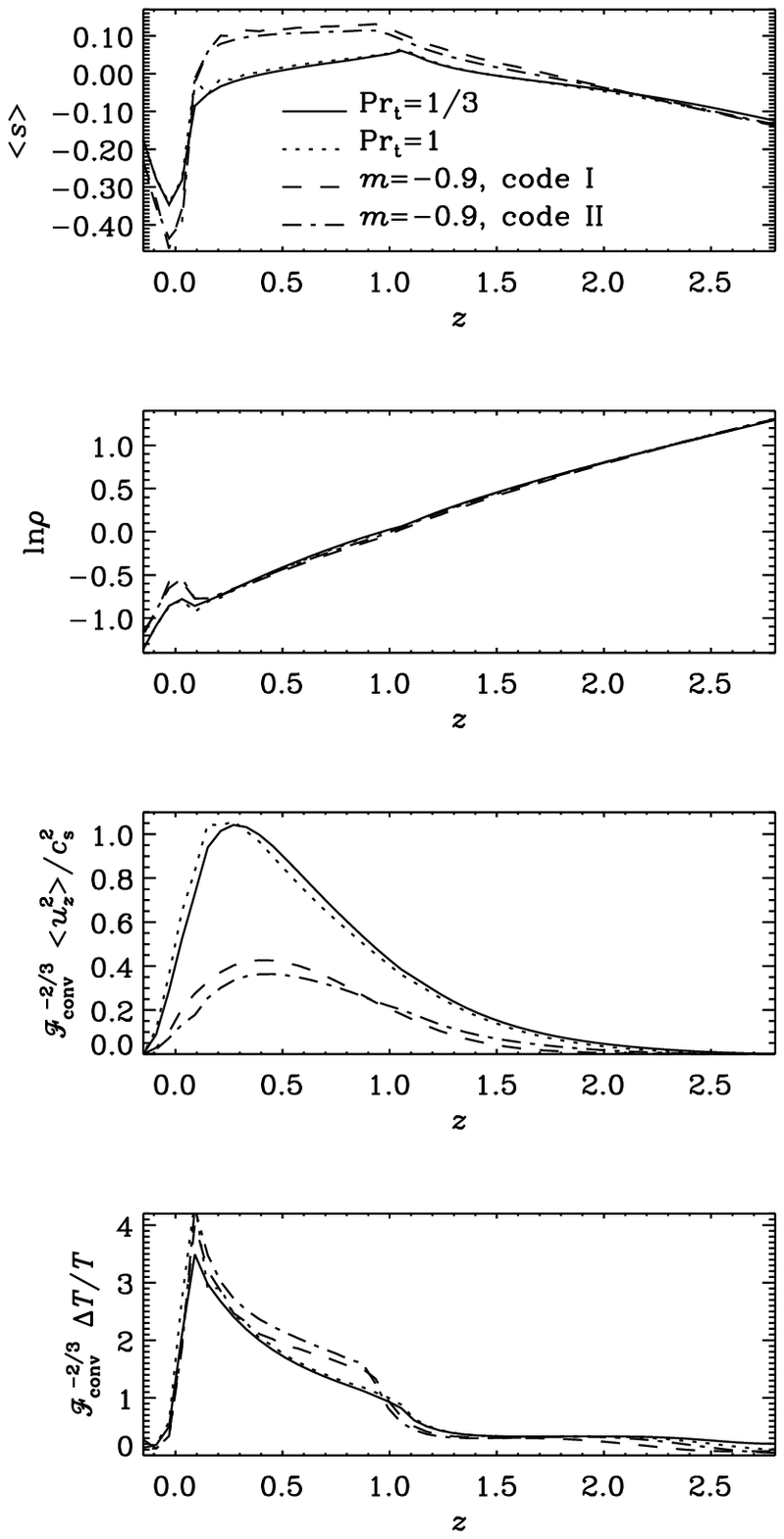}\caption[]{
Comparison of SGS models (solid and dotted lines, $m=-1$) with direct
calculations ($m=-0.9$).  The dashed and dash-dotted lines refer to 
Code I used in \Sec{Sres} (Nordlund \& Stein 1990) and Code II used in
the present section (Chan \& Sofia 1986).  While the $m=-1$ and
$m=-0.9$ models are very similar in the mean density stratification
and the scaled temperature fluctuations, there are significant
differences in the mean entropy profile and the scaled vertical
velocity fluctuations.  }\label{Fcomp_kwing}\end{figure}

Our next step is to compare our findings with the case where the
radiative flux is replaced by a Smagorinsky subgrid scale (SGS) energy
diffusion inside the convection zone (see Chan \& Sofia 1986, 1989),
corresponding to the limiting situation $m=-1$.  For this purpose, we
use the original code of Chan and Sofia.  Since the formulation
(solving the conservative form of the Navier-Stokes equations) and the
scheme of this code (second order only) are different from that
used in the previous section, we make a comparison of the results
of the two codes using the previous $m=-0.9$, $\nu=0.006$ case (with a
$50^3$ mesh).  In \Fig{Fcomp_kwing}, the dashed and dot-dashed curves
represent results from the Nordlund \& Stein (1990) code (I) and the
Chan \& Sofia code (II) respectively.  The agreement is good.

In the SGS case, the radiative conduction and cooling outside the
convective region are fixed in the same way as discussed in
\Sec{Seqn}.  Inside the convection zone, the numerical stability of 
the energy equation is maintained by a subgrid scale diffusive flux of
the form $=-\chi_{\rm t}\rho T\nabla s$, where $\chi_{\rm t}$ is the
SGS diffusivity.  
The numerical stability of the momentum equation is maintained by a
SGS kinematic viscosity
\begin{equation} \nu_{\rm t} = 0.32\,
\triangle x \triangle z\, (2 \mbox{\boldmath$\SSSS^2$})^{1/2}.
\end{equation}
where $\triangle x$ and $\triangle z$ are the horizontal
and vertical grid widths respectively.
The ratio between $\nu_{\rm t}$
and $\chi_{\rm t}$ is fixed throughout the convection zone.

\begin{figure}[t!]\includegraphics[width=\columnwidth]{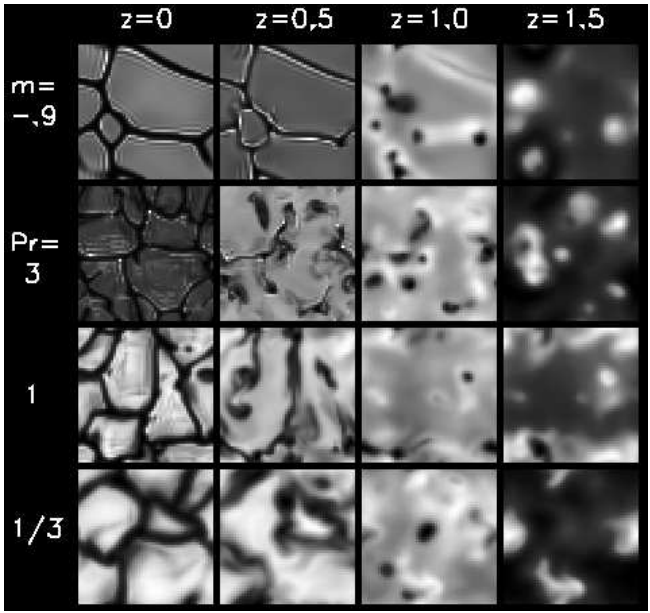}\caption[]{
Horizontal slice of temperature for $m=-0.9$ and SGS model using
three different values of the Prandtl number
$\mbox{Pr}=\nu_{\rm t}/\chi_{\rm t}$. ${\cal F}_{\rm conv}=0.01$.
All calculations here use a $50^3$ mesh.
}\label{Fkwing_tiles}\end{figure}

In the original code of Chan \& Sofia (1986) the effective Prandtl
number, ${\rm Pr}_{\rm t} \equiv \nu_{\rm t} /\chi_{\rm t}$, was
chosen to be 1/3.  However, in order to facilitate comparison with the
models discussed previously, where we varied the thermal diffusivity,
we now consider three models with ${\rm Pr}_{\rm t}=1/3$, 1, and 3.
All three models have the same total flux $F_{\rm tot}=0.01$ and use a
$50^3$ mesh.  In the SGS case the flow still shows the usual granular
pattern at the surface; see \Fig{Fkwing_tiles}.  Smaller diffusivity
$\chi_{\rm t}$ produces thinner intergranular lanes and decreases the
size of the smallest granular structures; it is consistent with the
trend shown in the previous section.  Compared to the constant $\nu$
case with the same mesh size (\Fig{FMMc2}), the SGS patterns show more
vortical features and indicate more vigorous turbulence.  In the case
with the largest $Pr$, however, the temperature field is getting a
little noisy; the thermal diffusivity is already too small to give
reliable results.  In \Fig{Fcomp_kwing} we plot the mean entropy,
density, as well as velocity and temperature fluctuations for ${\rm
Pr}_{\rm t}=1/3$ and $1$.

Since $\nu_{\rm t}$ depends on the local strength of the turbulence,
it is not uniform anymore.  Its horizontal mean varies with depth and
reaches a maximum near the top of the convection zone.  The range of
the mean values ($0.0015 - 0.0030$), however, is quite limited and
they are less than half of that used in the previous comparable case
($m=-0.9$, $\nu=0.006$ and same mesh size).  Relative to this case,
the effective Reynolds numbers of the SGS models are thus considerably
larger.  The SGS diffusivity $\chi_{\rm t}$, on the other hand, 
($\propto{\rm Pr}_{\rm t}^{-1}$ approximately) varies by almost an 
order of magnitude across the different models and are larger than the
$\chi$ of the low-resolution $m=-0.9$ case (ratio $\approx 1.1-10$).
The temperature fluctuations of the SGS models are thus smaller 
(difference $\approx 20-30\%$).  To deliver a comparable amount of 
convective energy flux, the value of the vertical rms velocity for the
SGS models is considerably higher than that of the $m=-0.9$ model.
This is accomplished by a somewhat steeper entropy gradient in the
convection zone.  The density stratification is similar.
Given that
the turbulence has become more vigorous, the kinetic energy flux is now
increased by almost a factor of 2 to about $35\%$ of the total flux.
Part of that ($\sim 9\% F_{\rm tot}$) is compensated by the subgrid scale
convective flux, and the rest is balanced by the enhanced enthalpy flux.

\section{Solar simulations}
\label{Srealist}

\begin{figure}[t!]\includegraphics[width=\columnwidth]{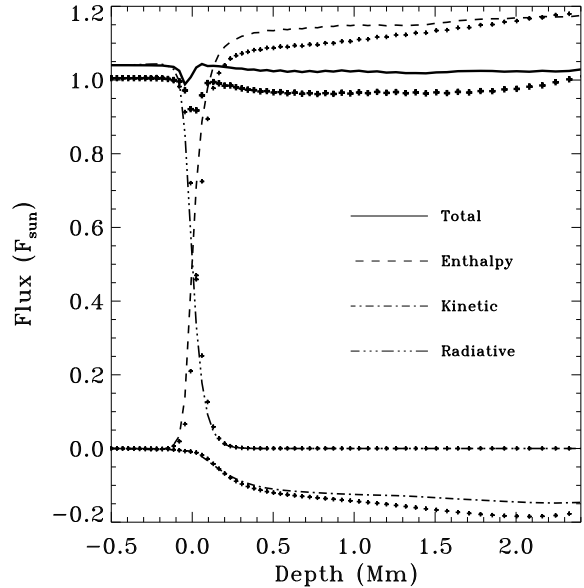}\caption[]{
Total, enthalpy, radiative and kinetic energy fluxes for the solar
simulations.  Crosses are case $\mbox{Pr}_{\rm t}=0.3$ and lines are case
$\mbox{Pr}_{\rm t}=3$.  When the diffusive energy flux is larger, the
kinetic energy flux increases and the enthalpy flux decreases slightly.
}\label{s_fluxes}\end{figure}

\begin{figure}[t!]\includegraphics[width=\columnwidth]{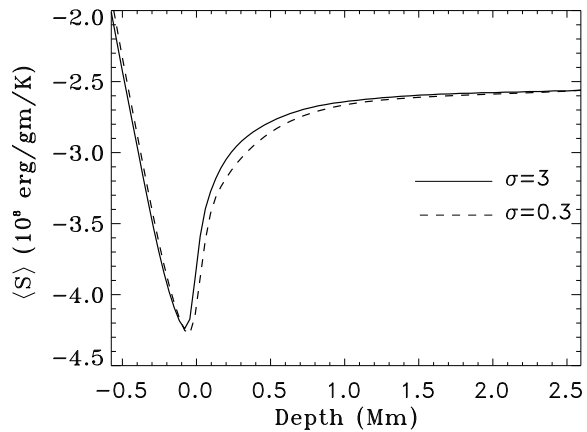}\caption[]{
Average entropy vs. depth.  (Here the Prandtl number is written as
$\sigma$.)  Larger energy diffusion produces a
slightly smaller entropy gradient just below the surface.
}\label{s_aventropy}\end{figure}

\begin{figure}[t!]\includegraphics[width=\columnwidth]{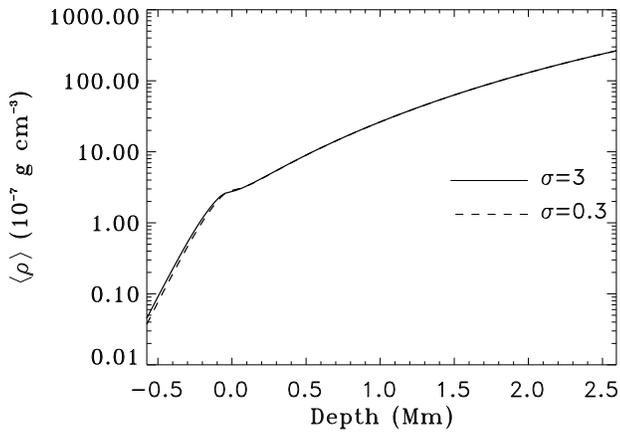}\caption[]{
Average density vs.\ depth.  Larger energy diffusion leads to a
less extended atmosphere.
}\label{s_avdensity}\end{figure}

\begin{figure}[t!]\includegraphics[width=\columnwidth]{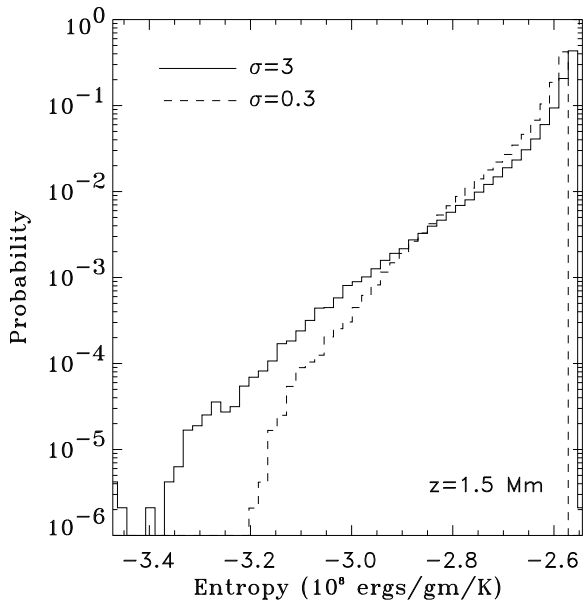}\caption[]{
Histogram of the entropy distribution at a depth of 1.5 Mm.  For
larger diffusive energy transfer, smaller Prandtl number, the lowest
entropy fluid is destroyed and the exponential distribution becomes
steeper.  
}\label{ss_hist_z1.5Mm_b}\end{figure}

\begin{figure}[t!]\includegraphics[width=\columnwidth]{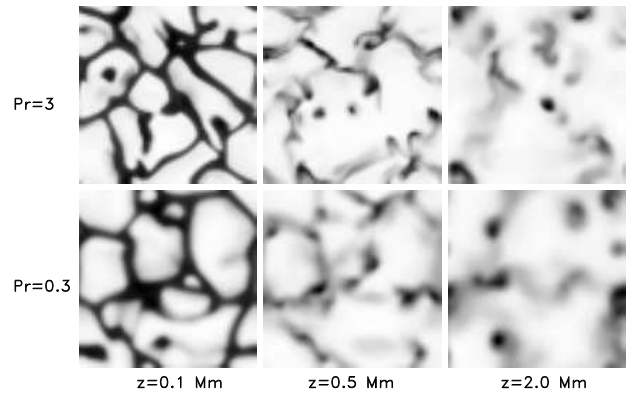}\caption[]{
Horizontal slices showing the temperature at depths of 0.1, 0.5 and 
2.0 Mm for two values of the Prandtl number.  Dark is low temperature
and light is high temperature.  Each panel is scaled independently.  
Smaller energy diffusion, larger Prandtl number, allows smaller scale
temperature structures to exist.
}\label{temp_tiles}\end{figure}

\begin{figure}[t!]\includegraphics[width=\columnwidth]{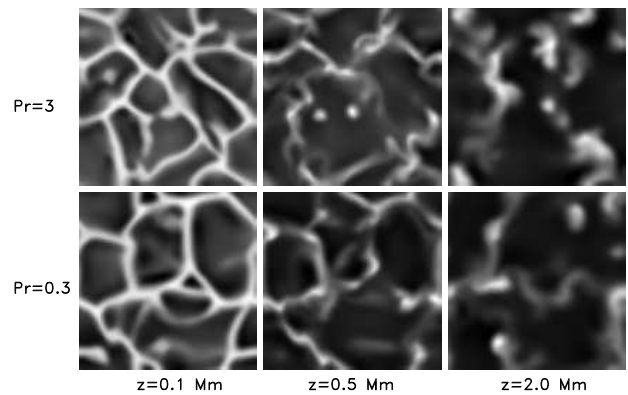}\caption[]{
Horizontal slices showing the vertical velocity fluctuations at
depths of 0.1, 0.5 and 2.0 Mm for two values of the Prandtl number.
Dark is upward velocity and light is downward velocity.  There is
little difference in the scale of the velocity structures between
these high and low energy diffusion cases.
}\label{velocity_tiles}\end{figure}

We have also made two simulations of near surface solar convection
with Prandtl numbers $\mbox{Pr}_{\rm t}=\nu_{\rm t}/\chi_{\rm t}=0.33$ and
$3.3$, using realistic physics, with a low resolution of $63^3$ (for details see
Nordlund \& Stein 1990; Stein \& Nordlund 1998).  Here the
radiative flux vanishes below the surface but there is numerical
energy diffusion with the above ratios to the momentum diffusion.
The vertical energy diffusion is proportional to the gradient of the
energy minus the horizontally averaged mean energy, so it is similar
to the entropy diffusion used in the SGS case.  The diffusion 
coefficients are not constant, but have terms proportional to the
sound speed, the magnitude of the velocity, and the compression.
They are enhanced where there are small scale velocity fluctuations
and quenched in laminar regions by the ratio of the magnitudes of the
third to the first derivatives of the velocity.  These simulations
only extend to 2.5 Mm below the surface and the convective flux is
controlled by specifying the entropy of the inflowing fluid at the
bottom.  The simulations were run for 2 solar hours.

Larger energy diffusion (i.e.\ smaller Prandtl number) produces a slightly
larger kinetic energy flux and a slightly smaller enthalpy flux,
resulting in a 5\% reduction in the net flux (\Fig{s_fluxes}).
The energy transport switchover between radiative and convective occurs
slightly deeper in the large energy diffusion case.

There are some other small alterations in the mean structure:  larger
energy diffusion produces a less steep mean entropy gradient at the
surface (\Fig{s_aventropy}) and a less extended atmosphere
(\Fig{s_avdensity}).

The low entropy fluid (which gives rise to the buoyancy work that
drives the convection) is fluid that reaches the surface and radiates
away its energy and entropy.  Larger energy diffusion destroys the
lowest entropy fluid as it descends back into the interior, by heating
it up.  This leads to slightly steeper exponential decline in the
entropy probability distribution function and a less extended low
entropy tail to the distribution (\Fig{ss_hist_z1.5Mm_b}).

\begin{figure}[t!]\includegraphics[width=\columnwidth]{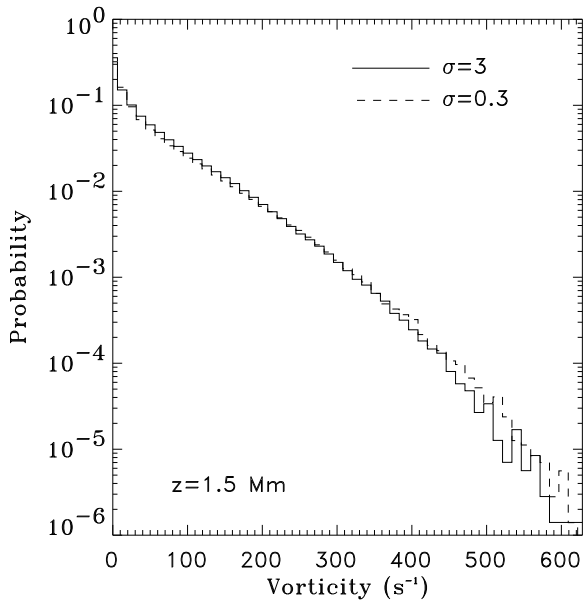}\caption[]{
Histogram of the vorticity distribution at a depth of 1.5 Mm.  Varying
the energy diffusion by a factor of ten has only a very slight effect
on the velocity and vorticity.
}\label{vorticity_hist}\end{figure}

Increasing the energy diffusion has a clear direct influence on the
temperature structures in these simulations, just as found when
varying the diffusive radiative flux -- larger energy diffusion
produces larger more diffuse temperature structures, smaller energy
diffusion allows smaller, sharper temperature structures
(\Fig{temp_tiles}).

The velocity and turbulence, on the other hand, are little affected by
this factor of ten variation in the energy diffusion
(\Fig{vorticity_hist}).  Changing the
resolution and hence the viscous momentum diffusion, however, has a
profound affect on the turbulence (vorticity) and the velocity (Stein
\& Nordlund 1998).  Less viscosity leads to greater turbulence,
larger vorticity, as well as a velocity distribution extending to larger
magnitudes in all directions but only in a small fraction of the
volume.

\section{Summary}
\label{Sconcl}

There are a number of clear changes in convective properties as the 
diffusive radiative flux is decreased while keeping the convective flux
constant in a convection simulation.
First, the entropy jump near the top becomes larger and steeper and 
the low entropy fluid produced by cooling at the surface penetrates
farther through the convection zone leading to
a finite probability to find small regions with very low entropy
near the bottom of the unstable layer. Second, the temperature
fluctuations increase and the velocity fluctuations decrease
in such a way that their product, which is proportional to the convective
flux, remains approximately constant.  Also, the kinetic energy
flux decreases.  Third, the dynamics in the
overshoot layer becomes somewhat more intermittent due to a few strong
downdraft plumes.  Finally, in all cases the
velocity and temperature fluctuations follow mixing length scaling laws;
see \Fig{Fprfluctn}.

The radiative flux really serves two different
purposes: it transports heat vertically, and it keeps the model numerically
stable by diffusing energy fluctuations both horizontally and
vertically.  Since those two properties appear to be reasonably
well decoupled from each other, one might separate them by
having a small vertical radiative flux plus a
subgrid scale diffusive flux that keeps the model stable. 
It is in practice difficult to decouple
the need for energy diffusion from the vertical diffusive heat transport,
which one would like to keep small if one diffuses on the temperature.
However, as discussed in the introduction,
such a separation is possible if the diffusive flux is based 
on entropy or temperature fluctuations.
Convective simulations with small radiative fluxes, as is appropriate
for cool stars, would then be feasible.
This was recently demonstrated by Miesch et al.\ (2000) and
Brun et al.\ (2004) using simulations of fully spherical shells.

\acknowledgements
AB thanks the Hong Kong University of Science and Technology for
hospitality. This work has been supported in part by the British
Council (JRS 98/39), the Research Grant Council of Hong Kong
(HKUST6081/98P), the Danish National Research Foundation through its
establishment of the Theoretical Astrophysics Center, and the PPARC
grant PPA/G/S/1997/00284, the National Science Foundation through grant
AST 0205500 and NASA through grants NAG 5-12450 and NNG04GB92G.

\end{document}